# ENTROPY PRODUCTION AND MORPHOLOGICAL TRANSITIONS IN NONEQUILIBRIUM PROCESSES

## Ph.D. Leonid M. Martyushev

### AUTHOR'S ABSTRACT OF DISSERTATION
for a Doctor of Science (Doctor habilitatus)
in physics and mathematics

**URAL FEDERAL UNIVERSITY**
Ekaterinburg
RUSSIA
2010

# GENERAL

**The objective of the Doctor's Dissertation** consists in a critical consideration of the maximum entropy production principle and an analysis of its corollaries for some (primarily morphological) nonequilibrium phase transitions.

In line with this objective, **three basic tasks had to be solved, as follows:**

1. A critical analysis of the approaches reported in the literature using variation principles based on the entropy production, including their generalization, classification, and proving.

2. Analytical and numerical studies of the initial stage of the morphological transitions and the coexistence phenomenon upon nonequilibrium crystallization. For this purpose, the notion of metastability and the maximum entropy production principle are used.

3. An experimental study of the initial stage of the morphological transition during a radial displacement of one fluid by another in a Hele-Shaw cell and a comparison with the analytical calculations, including those made on the basis of the entropy production computation.

**Novelty of the Doctor's Dissertation:**

1. It has been shown that the thermodynamic formulation of the maximum entropy production principle (MEPP) by Ziegler is similar to the microscopic formulation of the MEPP by Koller-Ziman. A generalized formulation of the maximum entropy production principle, which also holds for nonequilibrium phase transitions, has been proposed for the first time on the basis of the results of the dissertation research.

2. Two new thermodynamic arguments have been adduced to substantiate the maximum entropy production principle. They are based on the Onsager hypothesis (a nonequilibrium state is treated as a fluctuation) and the hypothesis of the invariance of the second law of thermodynamics in the transformation of the reference system for thermodynamic flows.

3. The concept of the entropy production as a criterion of selecting the morphological phases in the case of nonequilibrium processes has been developed.

3.1. The behavior of the entropy production in an arbitrary regime of the growth of a spherical and a cylindrical particle near the morphological transition has been studied for the first time. The maximum entropy production principle and a linear stability analysis have been used for the first time to construct a comprehensive morphological phase diagram (with stable, unstable, and metastable regions) for different regimes of the growth of a spherical and a cylindrical particle.

3.2. The entropy production and its variation upon a morphological transition in the problem of the radial displacement of a fluid in the Hele-Shaw cell have been derived. The maximum entropy production principle and a linear stability analysis of the displacement have been used for the first time to construct comprehensive morphological diagrams. The sequence of the morphological transitions depending on the displacement parameters has been specified.

4. The following has been done to substantiate the results mentioned in the item above:

4.1. A weakly nonlinear analysis of the morphological stability of a round particle in an arbitrary growth regime has been performed for the first time.

4.2. The initial stage of the loss of the morphological stability by growing round and spherical crystals (particles) has been analyzed numerically. The dependence of the critical size of the stability on the growth regime, the amplitude and the mode of perturbation has been established. It has been shown for the first time that, as the perturbation amplitude increases, the critical size of the stability always decreases to some value, which has been called "a binodal" in the dissertation.

4.3. The problem of the linear stability of the interface between fluids upon displacement at a constant flow rate in a radial Hele-Shaw cell has been solved for the first time taking into account all the factors determining the displacement process. An analytical expression for the critical radii of the loss of stability by the displacement front has been derived for all perturbation modes including the translation mode.

4.4. The critical radius of the loss of interface stability between air and silicon oil (PMS-5) depending on the thickness of the Hele-Shaw radial cell and the flow rate of the displacing fluid have been determined experimentally for the first time.

5. A possible relationship between the so-called $S$-shaped kinetic curves, the theory of extreme values, and the maximum entropy production principle has been noted for the first time.

6. The maximum entropy production principle has been used for the first time to predict the smallest Reynolds number, at which a transition from a laminar to a turbulent flow in a round pipe occurs.

**The basic statements of the Dissertation:**

1. The maximum entropy production principle has both thermodynamic and statistical foundation, and it can be generally formulated as follows: *at each level of description, with preset external constraints, the relationship between the cause and the response of a nonequilibrium system is established such as to maximize the entropy production*. In this formulation, the principle is also applicable to description of nonequilibrium phase transitions.

2. Prigogine's minimum entropy production principle and Ziegler's maximum entropy production do not contradict each other. The former is a corollary of the latter. It is inexpedient, and often erroneous, to generalize the local minimum entropy production principle to the integral case.

3. The prevalence of the $S$-shaped kinetic curves, which are observed during crystallization and other relaxation processes, does not contradict the maximum entropy production principle and can be understood with the help of this principle.

4. As the perturbation amplitude increases, the critical size of the morphological stability decreases in the case of nonequilibrium crystallization from some value - spinodal (the boundary of stability relative to infinitely small perturbations) to a minimum possible value, the so-called binodal. The morphological transition occurs in the metastable region (between the binodal and the spinodal); therewith the change of the crystal mass increases stepwise.

5. A necessary condition for the occurrence of a morphological transition is a larger entropy production in the final phase. The binodal of a morphological transition during nonequilibrium crystallization in a diffusion-limited growth can be found from the condition that the difference of the entropy productions in the first (unperturbed) and second (perturbed) growth regimes turns to zero.

6. The notion of the metastable region, which has been introduced for morphological transitions, allows explaining the experimentally observed phenomenon of the coexistence of different growth regimes.

7. The calculations of the entropy production in competing phases predicts the smallest Reynolds critical number, at which a transition from a laminar to a turbulent flow of a fluid can take place in a round pipe in the presence of arbitrary perturbations.

8. Modified boundary conditions in a linear analysis for the morphological stability and calculations of the entropy production allow explaining the experimentally observed translation instability during the radial displacement of a fluid in a Hele-Shaw cell, as well as the critical size of the stability of the displacement boundary shape in the presence of perturbations, which are not infinitely small.

**The scientific significance of the Dissertation** is determined by the potential of the proposed entropy production analysis for possible nonequilibrium transitions and, also, by the theoretical and experimental results of the work.

**The practical significance.** The results and the conclusions related to the maximum entropy production principle can be used to construct variational solutions of mathematical models of nonequilibrium processes and substantiate existing empirical kinetic regularities.

The results presented in the second part of the doctoral dissertation are important for production of crystals with special properties, since they determine the shape of the interface depending on thermophysical parameters, which control the process of nonequilibrium crystallization.

The results reported in the third part of the doctoral dissertation can be used to develop petroleum production technologies related to extraction of residual oil from wells. They can also be used to solve ecological problems connected with propagation of underground water and liquid pollutants in porous media.

**The personal contribution.** The author played the leading role in statement of the problem and research tasks, selection of basic pathways and methods of their solution, analysis and interpretation of the results, and preparation of all the published papers related to the doctoral thesis. All the analytical and numerical calculations, as well as experiments and their processing have been performed by the author jointly with the co-authors of the published papers.

**The organization of the Dissertation.** The doctoral thesis consists of an introduction, three parts, and a conclusion. The work has 268 pages, including 63 figures, 9 tables, and a list of 210 references.

# SUMMARY OF THE DISSERTATION

Formulated in **the Introduction** is the topicality of the research, the objective of the doctoral thesis, the scientific novelty, and the basic statements of the Dissertation.

**The first part of the doctoral thesis** is concerned with the foundations of the maximum and minimum entropy production principles. This part consists of two sections.

**The first section** is dedicated to a critical consideration of the entropy principles from the viewpoint of nonequilibrium thermodynamics. To start with, it briefly reports the fundamentals of linear nonequilibrium thermodynamics and variational principles by L. Onsager, I. Gyarmati, and I. Prigogine. Then the discussion is focused on a critical consideration and development of the variational principle by H. Ziegler (1957-1983) who proposed to find the explicit form of the dependence of thermodynamic flows $J_i$ on thermodynamic forces $X_i$ using the maximum entropy production principle: *If irreversible force $X_i$ is prescribed, the actual flux $J_i$, which satisfies the condition $\sigma(J_i) = \sum_i X_i J_i$, maximizes the entropy production*. In the mathematical form, the conditional maximum $\sigma(J_i)$ corresponds to

$$\delta_J \left[ \sigma(J_k) - \mu(\sigma(J_k) - \sum_i X_i J_i) \right]_X = 0, \qquad (1)$$

where $\mu$ is the Lagrange factor. Equation (1) easily gives a relationship between thermodynamic forces and flows, which generally is nonlinear:

$$X_i = \frac{\sigma(J_i)}{\sum_i \frac{\partial \sigma}{\partial J_i} J_i} \partial \sigma / \partial J_i. \qquad (2)$$

With the simplest choice of the entropy production in the form $\sigma = \sum_{i,k} R_{ik} J_i J_k$ (where $R_{ik}$ is a symmetrical tensor), the relationship (2) easily gives, as is shown in the doctoral thesis, all the basic relationships of local linear nonequilibrium thermodynamics. The doctoral dissertation describes a geometrical interpretation of Ziegler's principle and the background of this principle, which is connected with the theory of plasticity.

The interrelation between Ziegler's principle and the second law of thermodynamics is treated as a separate issue. A conclusion that the extremum (1) indeed is a maximum follows from the second law of thermodynamics. However, if the maximum entropy production principle is postulated first, the second law of thermodynamics can be derived as its corollary. Ziegler considered this point separately assuming a convexity of $\sigma(J_i)$ and one-to-one correspondence of the forces and the flows (2). However, even in the absence of these assumptions, the second law of thermodynamics can be derived as a corollary of the principle under discussion. Let in some imaginary system the entropy production *can* take the values smaller than zero at preset forces. Then, on the basis of the postulated principle, physically realizable flows will be such that the entropy production will be the largest; that is, the entropy production will be equal to the maximum *positive* number among the *possible* numbers. If it is assumed that the system cannot find flows satisfying the preset forces, so that the entropy production is greater than zero, then the system *always* has a variant, in which the values of the flows are taken to be zero. In this case, the entropy production will be zero, and this value will be a maximum in this bizarre example. Therefore, in line with the maximum entropy production principle, physically practicable states with a negative entropy production are never possible.

Attention is drawn in the dissertation to the possible "paradox" of the use of Ziegler's variational approach, in which thermodynamic forces (two or more) determined from Eq. (2) can differ from the initially chosen forces. This ambiguity of the expressions for thermodynamic forces is not a drawback of the Ziegler method, but it is rather a specific feature of all nonequilibrium thermodynamics, which is initially based on the balance equations for entropy, energy, momentum, and matter, as well as on the first two laws of thermodynamics. Interestingly enough, it is possible to find a condition, at which the set of forces is determined uniquely from the known entropy production. For example, for two forces it has the form

$$X_2(J_1,J_2)\frac{\partial \sigma(J_1,J_2)}{\partial J_1} = X_1(J_1,J_2)\frac{\partial \sigma(J_1,J_2)}{\partial J_2}, \qquad (3)$$

which, in the case of a quadratic function of the entropy production, is fulfilled if Onsager's reciprocal relations hold.

It has been shown in the doctoral dissertation how the Onsager variational principle can be derived from Ziegler's principle. Particular emphasis has been placed on well-known Prigogine's principle (the minimum entropy production principle). At first glance one can get a sensation that these two principles are absolutely contradictory to each other. But this is not so. Both linear and nonlinear nonequilibrium thermodynamics can be constructed deductively from Ziegler's principle. Prigogine's principle of the minimum entropy production follows already from linear thermodynamics as a particular statement, which is valid for stationary processes in the presence of some free forces. Thus, Prigogine's principle has a much narrower field of applicability than Ziegler's principle. These differences can be explained in a less formalized language as well. Assume a system with entropy production of a known type is considered. Then, if thermodynamic forces are preset, the system will adjust, in accordance with Ziegler's principle, its thermodynamic flows such that the entropy production is a maximum. If the entropy production is a quadratic function, this adjustment will lead to a linear relationship between the flows and the forces, with equal cross coefficients. Furthermore, if the system is in the stationary state, and some of the thermodynamic forces remain free, the flows formed by Ziegler's principle will decrease the thermodynamic forces, and the latter will, in turn, decrease the flows to the minimum entropy production. Thus, a hierarchy of the processes is observed: in a short time after a disturbance is introduced (on the order of time required for establishment of a local flow) the system maximizes the entropy production at the given fixed forces, and, consequently, linear relationships between the flows and the forces hold; then, in a time (on the order of the relaxation time of the free forces) the system decreases the entropy production to a minimum.

Both Ziegler's and Prigogine's principles are differential ones. However, there is a tendency in the literature to generalize the minimum production principle to the integral case. The Doctoral dissertation provides a detailed critical consideration to this issue by the example of the temperature distribution in a homogeneous rod on account of heat conductivity. The heat conductivity coefficient is assumed to be a $n$th power function of temperature (this does not contradict the range of applicability of Prigogine's principle). An analysis of the problem, which was performed in the Doctoral thesis, demonstrated that attempts to generalize Prigogine's local principle to the so-called integral case are unreasonable and, broadly speaking, erroneous. This is because even if additional restrictions are introduced as compared with the local formulation (on the relationship between the heat conductivity coefficient and temperature or the smallness of the temperature differential at boundaries of the system studied), the principle is invalid, probably except when the heat conductivity coefficient is inversely proportional to the temperature squared.

This section of the Doctoral dissertation gives much attention to methods substantiating the principle of the maximum entropy production, which Ziegler introduced as a postulate. Let us cite one of the possible approaches to thermodynamic substantiation of Ziegler's principle, which, in my opinion, can prove to be sufficiently interesting and promising if this approach is developed properly in a future. Assume that the second law of thermodynamics holds ($\sigma \geq 0$). Let $X = $ const $\geq 0$, and we have to prove that the system selects a maximum possible value of $J$ (and, hence, $\sigma = XJ$). Assume several different flows are possible. All of them should be larger than zero because $\sigma \geq 0$ (the flows are directed towards a decrease in the thermodynamic force). Suppose that the reference system of flows can be selected arbitrarily (e.g., by transforming the time scale). Among the possible flows, the maximum flow is chosen as the zero flow. Then, with respect to the selected reference system, all the other flows and the corresponding entropy productions will be negative. Since the second law is a universal law of nature and should not depend on such transformations (this can be viewed as one more postulate of the given proof, though), then we substantiate that at a preset force only the maximum possible flow and, hence, the maximum entropy production is implemented. It is particularly emphasized in the Doctoral dissertation that in the general case a rigorous substantiation of maximum entropy production principle in the context of just thermodynamic representations on the basis of sufficiently simple and intuitively physically clear postulates seems to be extremely difficult.

It should be acknowledged that H. Ziegler presented the material in a rather formalized manner and exemplified the use of the principle by just a few problems of the theory of plasticity and chemical kinetics, which can be solved by other alternative methods as well. This circumstance and, also, poor understanding of nonlinear phenomena and availability of other variational formulations in thermodynamics, which hold for a nonlinear case, were the reasons why little significance was attached to Ziegler's principle and it did not become public knowledge. However, as compared with other variational formulations of thermodynamics applicable to the nonlinear range, Ziegler's formulation is, in my opinion, most appropriate and simple. From the thermodynamic consideration of the maximum principle it follows a natural question of how this principle shows itself at the microscopic level. The answer to this question is dealt with in the next section of the first part of the Dissertation.

**The second section** discusses the maximum entropy production principle from the viewpoint of nonequilibrium statistical physics. The discussion in this section begins with a variational method used to solve the linearized Boltzmann equation for description of a rarefied gas. It has been shown that this well-known mathematical method of solution is reduced, thanks to the studies by M. Kohler (1948) and J. Ziman (1956), to the following statement: *in nonequilibrium gas systems the velocity distribution function is such that with preset temperature gradients, concentration, and average velocity, the density of the entropy production is a maximum on condition that it equals the sum of products of flows by forces* $\sum_k X_k J_k$. The mathematical notation of this statement reduces to the form

$$\delta_{\Phi,\mu}\left\{\sigma(\Phi) + \mu\left(\sigma(\Phi) - \sum_k X_k J_k(\Phi)\right)\right\}_X = 0, \qquad (3)$$

where $\Phi$ is a minor addition to the local Maxwell distribution function and $\mu$ is the Lagrange factor.

As can be seen, the mathematical notation of this principle is analogous to the notation of Ziegler's principle (1), but in (3) variation is not with respect to the flows $J_k$, as Ziegler did, but to the distribution function. Let us note one corollary of (3). It is known that at fixed forces the entropy production in a linear approximation is a function only of the kinetic coefficients $L_{ik}$: $\sigma = L_{ik} X_i X_k$. Therefore, finding the velocity distribution function, which maximizes the entropy production, we actually maximize the diagonal kinetic coefficients $L_{kk}$.

J. Ziman (1956) was the first who gave a thermodynamic interpretation of the variational principle (3) and ascribed it the status of a physical law, rather than just a mathematical method for solving the Boltzmann equation. He came up with an idea that Boltzmann's H-theorem is a kind of a proof of the second law of thermodynamics, and, by analogy, the variational theorem under consideration probably points to the existence of a sufficiently general statement as to the behavior of the entropy production in nonequilibrium systems (the maximum entropy production principle). At present this variational principle is assumed to be one of the main and efficient methods for solving the Boltzmann equation not only for classical gas systems, but, also, for electron and the phonon transport in solids. It should be noted that a similar variational method, which is connected with maximization of the entropy production, is used to find solutions of equations derived in terms of the linear response theory and valid for both quantum systems and sufficiently dense systems (H. Nakano, 1959-1960; D. Zubarev, 2002). This approach has also been discussed in the Dissertation.

The Doctoral dissertation critically reviews some papers (A. Filyukov, 1967-1968; W. Jones, 1983; R. Dewar, 2003, etc.) developing a hypothesis that the maximum entropy production principle is a generalization of the maximum entropy principle and suggests that not only the entropy of the state is a maximum upon establishment of equilibrium[1], but also the trajectory entropy upon establishment of a stationary nonequilibrium state is a maximum. It is quite possible that this statement indeed is true, but the attempts made to prove it have been, as is shown in the Dissertation, unconvincing so far[2]. The fact that maximization of the entropy production is connected with the most probable (effected by the largest number of microtrajectories) development of the system is seen from the following considerations. Assume, in line with L. Onsager (1931), that a nonequilibrium system near equilibrium (with the entropy $S_{eq}$) can be viewed as a fluctuation relative to the equilibrium state. Let at the moment of time $t_0$ an isolated system be brought to a nonequilibrium state with the entropy $S_0$. Assume that by the next moment of time $t$ ($t-t_0$ is much longer than the time of one collision, but is much shorter than the relaxation time) the system can pass to one of the states having the entropy $S_{-N},...,S_N$ (with $S_{-N}<...<S_0<...<S_N<S_{eq}$); since the process is spontaneous, part $S_i$ will be larger than $S_0$. The number of these states and the values of their entropy are naturally determined by the kinetic properties of the relaxing system, the initial state, and the time $t-t_0$. *On the other hand*, in accordance with the Onsager hypothesis, each of *new* states will be treated as some fluctuations relative to the equilibrium state. It is known that the probability of such fluctuations/states is proportional to $\exp(-(S_{eq}-S_i))$, and, correspondingly, the state with $S_N$ ($S_{eq}-S_N$ is a minimum) is most probable. However, $(S_N - S_0)/(t-t_0)$ is a maximum in this state. Since in an isolated system this value coincides with the entropy production, then it has been shown that the most probable will be an evolution of the system obeying the maximum entropy production principle.

Thus, the first part of the Dissertation presents available thermodynamic and statistical arguments for substantiation of MEPP. The demonstrated similarity of the nonequilibrium thermodynamic and linear kinetic formulations of the principle and the existing substantiations of these formulations predetermine considerable broadening of applications of this principle and pose the question as to further generalization of MEPP, particularly with respect to solution of problems related to selection of stable regimes in nonequilibrium processes. These issues for nonequilibrium crystallization and some hydrodynamic systems are considered in the next two parts of the Dissertation.

---

[1] Since the motion is unstable in a system of many particles, all possible microstates occur in the system, and the system of particles passes with time to a state with a maximum number of the microstates.
[2] In future the development of this idea can result in a unified approach used for consideration of equilibrium and nonequilibrium phenomena.

**The second part of the Dissertation** discusses morphological transitions during nonequilibrium crystallization and the phenomenon of simultaneous development of crystals of different shapes (coexistence) from the viewpoint of the maximum entropy production principle. This part consists of three sections.

**The first section** is concerned with experimental examples of the phenomenon of the loss of morphological stability and the phenomenon of coexistence of different morphological phases (see, e.g., Fig. 1).

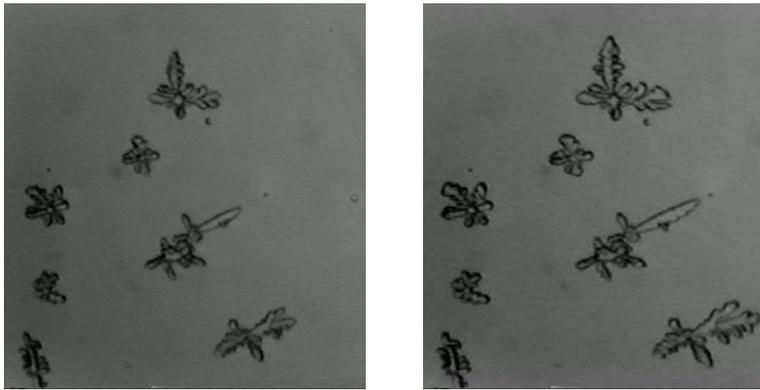

Fig. 1. Structures observed during crystallization of ammonium chloride from a water solution. Crystallization under room conditions of a solution initially saturated at 40°C. The visible region is 0.7 mm horizontally. Eighty seconds passed between the first and second fragments.

The Dissertation explains the reasons for distortions of the boundary, which are related to a concentration inhomogeneity of the solution. Some superficial resemblance between the transition from one form of the growth to another form (the morphological transition) and equilibrium phase transitions has been noted. However, two important questions arise. Does metastability occur upon transitions from one morphological phase to another? What is to be assumed an analog of the thermodynamic potential? The answer to these questions is the main topic of the second part of the Dissertation.

The morphological stability is usually studied with the following approximations (W. Mullins, R. Sekerka, 1963)[3]: quasistationarity of the process (valid at relatively small supersaturation); isotropy of the kinetic coefficient of crystallization and surface tension, and the initially simplest (e.g., cylindrical or spherical) shape of the crystal. The mathematical representation of the problem is as follows:

$$\Delta C = 0, \qquad (4)$$

$$D\frac{\partial C}{\partial \vec{e}}\bigg|_r = \beta(C_{\text{int}} - C_{\text{int }eq})\bigg|_r, \qquad (5)$$

$$C(r_\infty) = C_\infty, \qquad (6)$$

---

[3] It is only in this approximation that the analytical solutions can be advanced sufficiently far.

$$V = \frac{D}{C_{sol} - C_{int}} \left.\frac{\partial C}{\partial \vec{e}}\right|_r, \qquad (7)$$

where $\beta$ is the kinetic coefficient of crystallization; $C$ is the concentration in the solution; $C_\infty$ and $C_{int}$ are the concentrations of the dissolved substance far from the crystal and at its surface; $C_{int\ eq}$ is the equilibrium concentration of the dissolved substance near the surface; $C_{sol}$ is the density of the crystal; $r$ is the equation of the crystal surface; $\vec{e}$ is the normal to the surface described by the radius-vector $r$; $D$ is the diffusion coefficient; $r_\infty$ is the radial coordinate far from the crystal; $V$ is the local velocity of the crystal.

Stability investigations are traditionally performed with perturbations of the initial growth shape by single harmonics of an amplitude $\delta$, and equations (4)-(7) are solved to determine the minimum critical size of the crystal, at which the perturbation amplitude begins to grow. Thus, the critical size of a crystal is the main characteristic of a morphological transition.

If we assume that the perturbation amplitude is infinitely small and limit ourselves to the linear perturbation theory, the critical sizes of stability are known for the growth of a spherical $R_S^S$ and a round $R_C^S$ crystal (W. Mullins, 1963; S. Coriell, 1965). These sizes, which are rendered dimensionless to the critical radius of nucleation, are as follows:

$$R_S^S = 0.5\ (1\ +\ 0.5(l+1)(l+2))\left[1 + \sqrt{1 + 2\alpha_1 \frac{(l+1)(l+2)}{(1+0.5(l+1)(l+2))^2}}\right], \qquad (8)$$

$$R_C^S = \frac{1 + A_\lambda k(k+1) + \sqrt{(1 + A_\lambda k(k+1))^2 + 4\alpha_2 k(k+1)}}{2}, \qquad (9)$$

where $\alpha_1 = D/\beta R_S^*$ and $\alpha_2 = D/\beta R_C^*$ are dimensionless complexes characterizing the growth regime (the growth is limited by diffusion at small values, and by surface phenomena at large values); $A_\lambda$ is a dimensionless complex related to $r_\infty$; $l$ and $k$ are numbers of perturbing modes; $R_S^*$ and $R_C^*$ are critical radii of nucleation of spherical and cylindrical shapes.

Formulas (8)-(9) fully determine stability of growing spherical and round particles with respect to infinitely small perturbations. Obviously, using the terminology available in physics of equilibrium phase transitions, these sizes can be thought of as spinodals of morphological transitions.

The classical linear analysis says nothing about what will happen if the perturbation amplitude is not infinitely small. To answer this question, the Dissertation included a weakly linear analysis (to the third order in $\delta$) of the problem (4)-(7) for a round crystal. It was found that at each harmonic the critical size of stability decreases as the perturbation amplitude grows (and, in some cases, comes to saturation). The results of a weakly linear analysis hold for relatively small perturbation amplitudes. What the behavior of the critical size of stability will be when perturbations increase unlimitedly can be understood only from a numerical solution of the problem (4)-(7).

The first section of the second part of the Dissertation critically discusses the literature data on the use of the maximum entropy production principle for description of morphological (nonequilibrium) transitions and the possibility to use the entropy production as an analog of the thermodynamic potential, which determines stability of equilibrium phases. The fundamental papers on this topic were written by D. Temkin, 1960; Y. Sawada 1983-1984; J. Kirkaldy, 1984-1995; E. Ben-Jacob, 1989-1990; and A. Hill, 1990. From the analysis of the literature data and the results discussed in the first part of the thesis, it was possible to advance the following hypothesis: *a necessary condition for a morphological transition is a larger entropy production in the final nonequilibrium phase, and an equality of the entropy productions of two nonequilibrium phases determines the binodal of the transition*[4].

In closing this section of the Dissertation discusses several experimental results in the context of the introduced hypothesis and the maximum entropy production principle. Only two results will be mentioned here. *First*, a consideration has been given to a stepwise (as regards the crystal growth rate) morphological transition from a stable needle-like morphology of a crystal to a platelet growth, which is observed in experiments on nonequilibrium solidification of a supercooled film of bidistilled water (A. Shibkov, 2003). It was found that this transition takes place at supercooling of 7.5 degrees. According to the reported data, in the vicinity of the transition point the dependence of the growth rate $V$ on supercooling $\Delta T$ is described well by a linear function $V=L(\Delta T-\theta)$, where $L$ and $\theta$ are some empirically determined dimensional coefficients. The values of these coefficients ($L$ and $\theta$) are (0.31 cm/(ºC·s), 3.5ºC) for a stable needle and (0.78 cm/(ºC·s), 5.0ºC) for a platelet, respectively. Using these data, the entropy production for each of the structures can be written as proportional to $L(\Delta T-\theta)^2$. The calculations show that in the range of existence of the crystalline structures at hand the entropy productions of the needle and the platelet prove to be equal exactly at supercooling of 7.5 degrees. *The second example* refers to *S*-shaped kinetic curves showing the dependence of the fraction of the solidified phase $q$ on time $t$, which are often observed, e.g., during mass crystallization. These unity-normalized experimental kinetic curves are described well by the model $q(t)=1-exp(-ct^n)$ ($c$ and $n$ being some parameters larger than zero). Crystal nuclei occur randomly in space, and, therefore, the time $T$, at which a transition takes place on an arbitrary site, can be assumed to be a random value and $q(t)$ can be thought of as the probability that a random site of the solution solidifies by the moment of time $t$. Therefore, in this case, $q(t)$ is a distribution function, and, on account of its explicit form, it is the Weibull distribution. However, it is known from mathematical statistics (E. Gumbel, 1962) that this type of distribution corresponds to the distribution of minima of random values. That is, if $T = \min\{T_1, T_2, ..., T_n\}$ and $n\to\infty$ ($T_1, ..., T_n$ being random identically distributed values, which are not less than zero), the distribution function $T$ is the Weibull function. Then it can be proposed that the transition at hand is complete in a minimum possible time or, analogously, crystallization progresses at a maximum possible rate. Since the crystallization rate is directly proportional to the entropy production in a system, a conclusion can be made as to maximization of the entropy production. The proposed approach, which relates three phenomena (the $S$-shape of the kinetic curve, the distribution of extreme values, and the maximum principle), provides a unified view of the numerous facts concerning nonequilibrium kinetics and relaxation reported in the literature.

---

[4] In line with the terminology used in the theory of equilibrium phase transitions, the binodal is assumed to be a boundary separating the region, in which the phase is stable, from the region, in which it is metastable and unstable.

**The second section** presents the results of numerical investigations on morphological stability of the crystal surface relative to perturbations of an arbitrary amplitude. The calculations were performed using the finite element method in special toolboxes of the MATLAB software. The set of equations (4)-(7) was solved numerically for a growing round and spherical crystal. Cosine functions in the first case and axial-symmetric spherical functions in the second case were imposed as perturbations. The results of linear and weakly linear analyses were used for testing the calculation algorithm at low-amplitude perturbations. The calculated dependences of the critical sizes $R_c$ on the perturbation amplitudes $\delta$ for the two- (see Fig. 2) and three-dimensional problems were qualitatively similar.

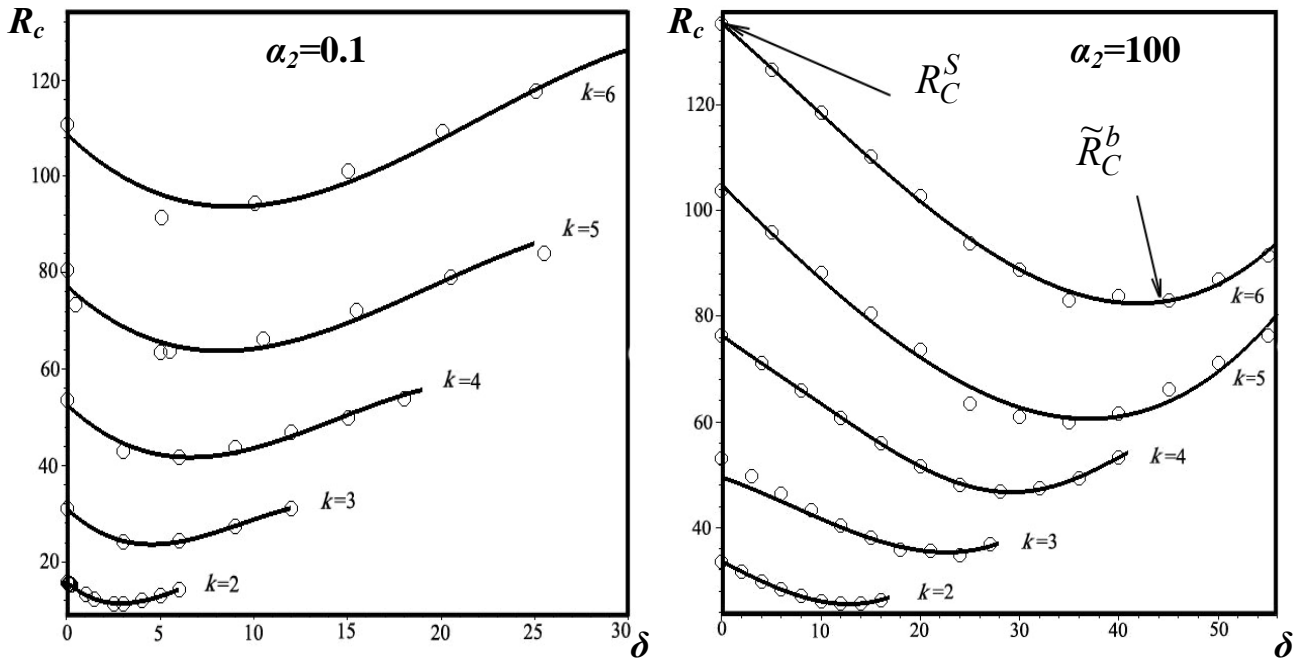

Fig. 2. Dependence of the critical size of the morphological stability $R_c$ on the perturbation amplitude $\delta$ in different regimes of the growth $\alpha$ and different perturbation modes $k$. The points indicate results of the numerical calculation; the solid line denotes results of cubic polynomial interpolation.

It was found that in any (from diffusion-limited to kinetic) growth regime and at any modes of the initial harmonic perturbations the dependences of $R_c$ on $\delta$ are similar and have two characteristic specific points: at $\delta \to 0$ and a minimum point.

While the first point ($R_C^S$ in Fig. 2) was studied well in analytical terms by methods of classical linear analysis for stability, the presence of the second point ($\widetilde{R}_c^b$ in Fig. 2) is a rather interesting and nontrivial result. Let there be a growing particle of a round shape. Assume that it develops in a medium with perturbations of a certain frequency and an *arbitrary* amplitude. Then, in accordance with the calculations (Fig. 2), the transition from a stable to an unstable growth takes place at some critical size corresponding to the minimum in the dependence of $R_c$ on $\delta$. If the experiment is carried out more "accurately" (that is, the level (the amplitude) of perturbations in the medium is reduced below some critical value), the size of instability will increase in accordance with the calculations. In line with the terminology accepted in the theory of equilibrium phase transitions, it is reasonable if the calculated minimum critical size $\widetilde{R}_C^b$ is called, by analogy, a binodal, and the stability size $R_C^S$, which was observed when the perturbation amplitude was almost zero, is termed a spinodal of a nonequilibrium transition. Thus, the transition to an unstable growth can be observed in the range from $\widetilde{R}_C^b$ to $R_C^S$ depending on the amplitude of perturbations. This range will be referred to as a metastable region by analogy with the theory of equilibrium phase transitions.

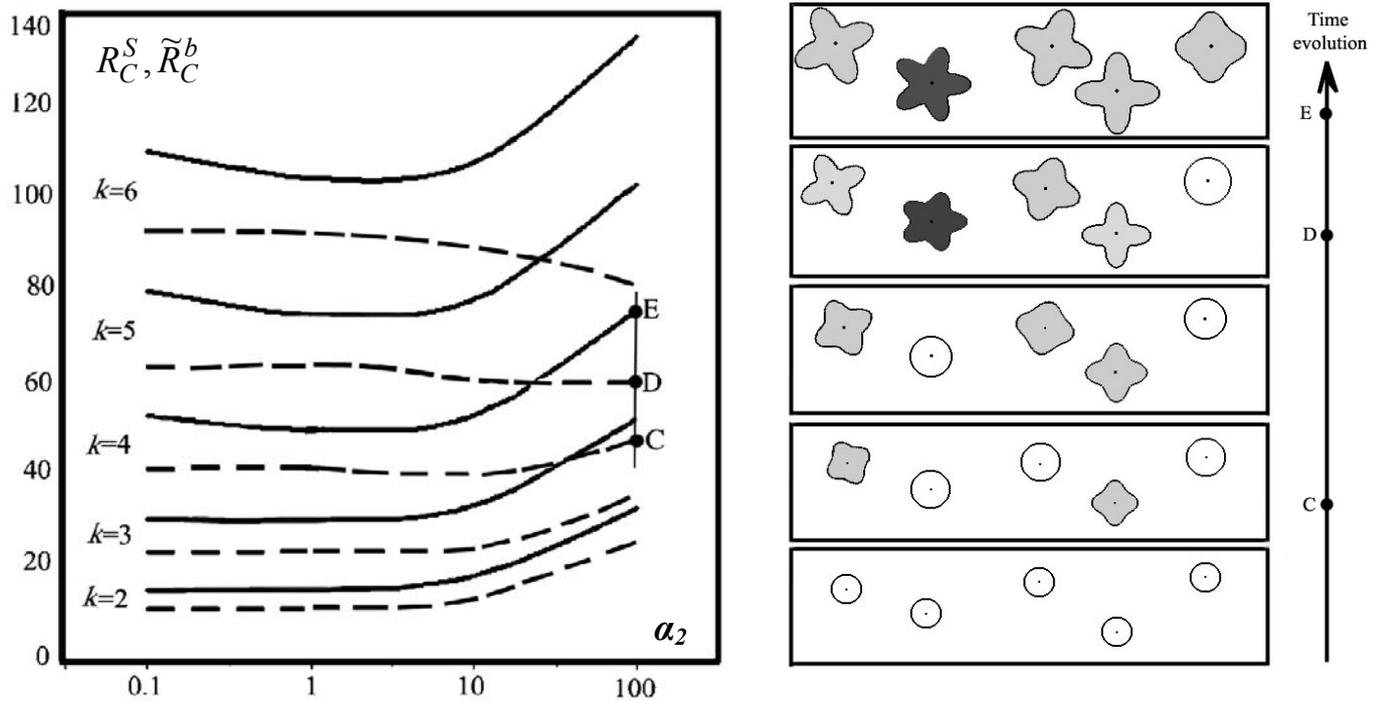

Fig. 3. The plot on the left shows the dependence of the spinodal $R_S$ (the solid line) and the binodal $\widetilde{R}_C^b$ (the dashed line) on the growth regime $\alpha$ at different perturbation modes $k$. The figure on the right shows the possible time evolution of growing particles in a medium (the growth conditions correspond to the trajectory CDE).

Figure 3 shows the behavior of the metastable region depending on the growth regime at different harmonics. In the region α < 1, which corresponds to the diffusion-limited growth regime, the metastable regions corresponding to different harmonics do not intersect. In the transition region (1<α<10) the binodal of the harmonic $k+1$ and the spinodal of the harmonic $k$ approach each other, and at α > 10 they intersect. As a result, the metastable regions of the critical radii overlap at adjacent harmonics. If *α* > 100 (exceptionally the kinetic growth regime), three or more metastable regions can intersect. Therefore, in the intermediate and kinetic regimes of growth in a medium with perturbations of different amplitudes and frequencies a large number of particles having different shapes — different morphological phases — can coexist and develop from round nuclei.

Illustrate the foregoing by a particular example. Let a multitude of round particles grow in a medium with fluctuations at $k \geq 4$, which have an arbitrary amplitude and lead to distortions of the boundary. We shall assume that the larger the amplitude of perturbations, the more rarely they occur, and, on the contrary, perturbations with a vanishingly small amplitude are infinitely numerous. Let the physicochemical parameters of the medium and a particle formed therein correspond to $\alpha_2 = 100$. Then the growth of the particles follows a straight line CE (see Fig. 3). Up to the point C all the particles have a round shape. In the interval CD particles of two types can be observed simultaneously: round particles (which are more probable considering the above assumptions of the perturbation statistics) and particles that have lost stability with respect to perturbations with $k = 4$. Beyond the point D (the binodal for perturbations with $k = 5$) the third type of particles can appear in the medium on account of the loss of stability by round particles under the action of perturbations with $k = 5$. Thus, three morphological phases coexist within the interval DE. Beyond the point E (the spinodal for perturbations with $k = 4$) all the remaining round particles become unstable with respect to perturbations with $k = 4$.

Thus, the numerical calculations of the morphological stability of growing round and spherical particles revealed the presence of metastable regions, pointing to the possibility that different forms of crystal nuclei can appear and grow simultaneously.

**The third section** deals with the possibility to use calculations of the entropy production and the hypothesis proposed above for quantitative prediction of the numerically calculated binodals.

As is shown in the Dissertation, the local entropy production $\Sigma_S$ in an element of the solution volume near the crystal surface can be written as

$$\Sigma_S \sim V^2 d\Omega, \tag{10}$$

where $V$ is the local velocity of the crystal and $d\Omega$ is an element of the solution volume near the crystal surface.

With the proposed hypothesis, the solution of $\Delta\Sigma_S = 0$ for the crystal size $R$ allows finding the transition binodal (here $\Delta\Sigma_S$ is the difference between the local entropy productions near the surface of two morphological phases[5]). The solution of this equation for spherical and cylindrical (particularly round) geometries in the case of the mathematical model (4)-(7) gave the binodals of spherical $R_S^b$ and round $R_C^b$ crystals of the form[6]:

$$R_S^b = \frac{l^3 + 2l^2 + l - 2 - 2\alpha_1(l+1) + \sqrt{(l^3 + 2l^2 + l - 2)^2 + 4\alpha_1^2(l+1)^2 + 4\alpha_1(l^4 + l^3 - l^2 + l + 2)}}{4l}, \quad (11)$$

$$R_C^b = \frac{1}{2}\left\{1 - \frac{\alpha_2 k}{2k-1} + \frac{2A_\lambda k(k^2-1)}{2k-1} + \sqrt{\left[1 - \frac{\alpha_2 k}{2k-1} + \frac{2A_\lambda k(k^2-1)}{2k-1}\right]^2 + 4\alpha_2 \frac{k(2k^2-1)}{2k-1}}\right\}. \quad (12)$$

---

[5] In this case, one morphological phase implies the initial (spherical or round) form of growth, and the other phase means the initial form with some added harmonic.

Using (8), (9), (11), and (12), full morphological phase diagrams of the regions of stable, metastable, and unstable growth of spherical and cylindrical crystals were constructed in the Dissertation. In a wide interval of the parameters the metastable regions corresponding to different perturbing harmonics overlap (similar to those determined numerically, see, e.g., Fig. 3), pointing to the possibility that a great number of morphological phases can coexist. It was found that the change in the mass of a crystal increases stepwise upon the morphological transitions. The value of the stepwise change decreases as the kinetic crystallization coefficient diminishes; the relative supersaturation decreases; the surface tension coefficient increases; and the numbers of perturbing harmonics grow.

Табл. 1.
Table 1. Binodal radii determined numerically and analytically (12) for a round crystal

| $\alpha_2$ | $k$ | $\widetilde{R}_C^b$ | $R_C^b$ |
|---|---|---|---|
| 0.1 | 2 | 11.3 | 10.8 |
|  | 3 | 23.6 | 24.6 |
|  | 4 | 41.7 | 43.2 |
|  | 5 | 63.8 | 66.6 |
|  | 6 | 93.7 | 94.9 |
| 1 | 2 | 11.8 | 10.6 |
|  | 3 | 23.9 | 24.4 |
|  | 4 | 42.0 | 43.0 |
|  | 5 | 63.8 | 66.5 |
|  | 6 | 92.5 | 94.5 |
| 10 | 2 | 14.2 | 9.2 |
|  | 3 | 25.1 | 23.0 |
|  | 4 | 40.9 | 41.7 |
|  | 5 | 60.8 | 65.2 |
|  | 6 | 89.3 | 93.6 |

The most important question is to what extent the binodal radius, which was calculated using the entropy production, coincides with its numerically predicted counterpart. A quantitative comparison of the results obtained for a round crystal[7] is given in Table 1. Whatever the perturbation mode, the accuracy of prediction based on (12) for the diffusion and intermediate regimes of growth was very high, with the discrepancy being just 2 to 10 percent. However, for the kinetic growth regime, the coincidence was much worse at perturbations with small $k$ (a maximum discrepancy as large as 35 percent at $k = 2$). A possible explanation of the observed discrepancy can be an insufficient accuracy of the analytical calculations performed for these range of the parameters. Let me explain the last point. Analytical calculations are based on comparison of the entropy production in a solution at the surface of unperturbed and perturbed growing particles. The entropy production is essentially a measure of nonequilibrium, but nonequilibrium is extremely small for the region at hand. Indeed, the farther we are from the diffusion growth regime ($\alpha_2$ increases), the more homogeneous is the diffusion field at the surface of a particle. Also, the more long-wave is the perturbation of the boundary, the closer is the curvature and, hence, the equilibrium concentration of such a perturbed particle to its unperturbed value. For these reasons, both the absolute value of the entropy production and the difference of the entropy productions for perturbed and unperturbed forms of the crystal growth are very small, and this can be a reason for an inaccurate analytical result for kinetic growth regime.

---

[6] These radii were rendered dimensionless to the critical radius of nucleation.
[7] The results were similar for a spherical crystal.

Thus, a quantitative coincidence has been revealed for the binodals predicted analytically using the entropy production and those calculated numerically when there is a relatively large gradient of the concentrations near the surface of a growing particle.

**The third part of the Dissertation** discusses the applicability of the maximum entropy production principle to hydrodynamic systems and nonequilibrium transitions therein. This part consists of three sections.

**The first section** is dedicated to a critical review of two large independently developing trends of research connected with the use of the maximum entropy production principle in hydrodynamics. The first of these trends was initiated by G. Paltridge (1975-2001). He calculated average annual climate on the assumption that the thermodynamic dissipation, which is due to horizontal flows of energy in atmosphere and oceans, is a maximum (in other words, among the multitude of stable states, a system will choose the one, in which maximum entropy is produced). As a result, using a sufficiently simple model, Paltridge found global average annual distributions of temperature, heat flows, and cloudiness on the Earth, which agreed well with the observed characteristics. Therefore, Paltridge's method involving the maximum entropy production principle become popular in studies concerned with climate on the Earth and other planets of the solar system.

In the second trend, the maximum entropy production principle was introduced axiomatically for calculating the time evolution of velocity fields at developed turbulent flows (R. Robert, J. Sommeria, 1992). A remarkable feature of the derived equations is that, while accounting for a small-scale motion and smoothing it, they satisfy the energy conservation law and other integrals of motion. A comparison of such calculations with direct numerical computations is in favor of the introduced principle.

This section ends with the following conclusions: 1) MEPP appeared in hydrodynamics intuitively for the most part, and it is substantiated primarily by its successful use. Extremely cumbersome computations, which are performed currently using MEPP in hydrodynamics, make its comprehensive verification difficult and hamper detection of possible limitations. 2) MEPP was not practically used in studies of nonequilibrium phase transitions in hydrodynamic systems, specifically investigations on instability of the shape of a moving phase boundary.

**The second section** discusses a classical transition from a laminar to a turbulent motion in a smooth round pipe. This case of the fluid flow under the action of a pressure gradient has received the most comprehensive experimental and theoretical study. It is known that this transition usually occurs at critical Reynolds numbers ($Re_c$) of about 2300. However, if an effort is made to reduce perturbations of the fluid flow under study (e.g., at the pipe inlet), the transition to the turbulent motion can be considerably postponed ($Re_c$ increases to $10^5$ or larger). It was shown analytically that the flow at hand is linearly stable at any $Re$. Therefore, the *upper* limit of the transition from a laminar to a turbulent regime probably does not exist, and, thus, the spinodal of this transition is infinite. However, an interesting question arises whether the *lower* critical Reynolds number exists. Obviously, to reach this number, a laminar flow of fluid should be exposed to perturbations. It was shown experimentally that the smallest number $Re_c$ is about 1760 for short-wave perturbations (A. Darbyshire, 1993) and about 1200 for long-wave perturbations (B. Benhamou, 2004). It can be concluded therefore that the experimentally found binodal of the transition from a laminar to a turbulent flow is 1200 in the presence of perturbations having arbitrary frequencies. One more indirect proof of this value is provided by recent numerical calculations of the flow in a round pipe (H. Faisst, 2003; H. Wedin, 2004). They revealed three-dimensional structures, the so-called "travelling waves", which have different azimuthal symmetries and represent, in the authors' opinion, the first manifestation of the transition to turbulence. The most interesting point is that 1250 is the smallest Reynolds number, at which these structures appear.

Let us make an attempt to predict theoretically the binodal value. In accordance with the approach developed in the Dissertation, it is necessary to equate the entropy productions of each (laminar and turbulent) of the nonequilibrium phases. When a fluid moves in a pipe, the entropy production is directly connected with dissipation of the mechanical energy produced by the pressure drop Δp at the ends of the pipe. If the temperature and the density of the fluid are constant, it can be assumed that the entropy production σ of the flow at hand is directly connected with the so-called friction (resistance) factor λ$(Re)$ (A. Reynolds, 1979; I. Idelchik, 1992), namely, σ$(Re)$~Δp·$Re$~λ$(Re)$·$Re^3$. Therefore, an analysis of the variation of the entropy production near a nonequilibrium transition at a preset $Re$ can be replaced by an analysis of *λ*. For a laminar flow, the resistance law (Hagen-Poiseuille law) has the form λ=64/$Re$; for a turbulent flow, the simplest and efficient is the Blasius law λ=0.316/($Re^{0.25}$). Upon transition from a laminar to a turbulent flow, the friction factor (and, hence, the entropy production) changes abruptly from small values (related to the Hagen-Poiseuille curve) to large values (related to the Blasius curve). This jump is the larger, the farther to the metastable region (larger $Re$) we "penetrate" in experiment by reducing the perturbations of the laminar flow. The Reynolds number, at which the entropy productions (the friction factors) are equal for the laminar and turbulent phases, is just 1200. Thus, it has been demonstrated in this section that the proposed theoretical method, which is based on calculation of the entropy production, can provide important quantitatively verifiable information about one of the most extensively studied nonequilibrium transitions, namely, the transition from a laminar to a turbulent flow in a round pipe.

**The third section** is concerned with experimental and theoretical studies of the loss of stability of the interface between two fluids during displacement in a radial Hele-Shaw cell. In this process, a less viscous fluid displaces a more viscous fluid while moving horizontally in a radial Hele-Shaw cell (two plane-parallel plates spaced a short distance from each other; the displacing fluid is injected into the cell in its center) and, as it moves, at some moment of time the initially round interface of the two fluids loses its stability and gets distorted, transforming to an intricate "fingered" structure.

This section begins with noting the value of this system for the range of problems discussed in the Dissertation, primarily the possibility of performing not only qualitative, but also quantitative experimental verification of the calculated size of morphological stability[8]. An analysis of the available experimental investigations into stability of the displacement front in a radial Hele-Shaw cell led to a conclusion that most research dealt with the stage of a supercritical growth of the fingers and their branching, as well as the effect of various artificial disturbances of the azimuthal symmetry of the cell on the morphology of the fluid interface after the loss of stability. The critical radius of the loss of morphological stability was measured in none of the experimental studies dedicated to the displacement in a Hele-Shaw cell. However, as is shown in the second part of the Dissertation, it is this information that is most important for the problems related to quantitative verification of conclusions based on the above-proposed hypothesis.

Then this section continues with experimental studies on the critical size of morphological stability taking the "air – silicon oil (PMS-5)" system as an example. The experimental setup, the experimental techniques, image analysis, and the procedure of finding the size of morphological stability have been described.

In the experiments, the critical size of instability ($R_{ex}$) was determined for each surface harmonic mode *n* (including the first, the so-called translational, mode) at several values of the incoming air flow rate ($Q$) and the distance between the glasses (*b*). (Table 2 gives only *the smallest* values of the measured critical size of stability). In some cases, the critical sizes corresponding to different developing modes proved to be nearly equal, and, as a result, several modes developed simultaneously (the so-called coexistence was observed). The measured critical sizes corresponded to some small perturbations, which were present in the experiment. Clearly the question arises as to how the critical size of morphological stability will change if large perturbations are introduced purposefully. This problem was studied with respect to translation instability caused by the shear action on the top glass of the Hele-Shaw cell. This study revealed a decrease in the critical size of instability relative to the first-mode perturbation.

---

[8] This verification of morphological stability during crystallization was practically impossible.

Table 2. Summary table of measurement and calculation results

| $Q$, mm²/s | $b$, mm | Number of experiments | $n$ | $R_{ex}$, mm | $R_S$, mm | $R_b$, mm |
|---|---|---|---|---|---|---|
| 123.0 ± 0.4 | 0.60 ± 0.03 | 7 | 1 | 18.2 ± 1.2 | 17.2 | 5.5 |
|  |  |  | 2 | 16.8 ± 1.1 | 30.3 | 17.5 |
| 226.6 ± 1.2 | 0.60 ± 0.03 | 4 | 2 | 15.2 ± 0.9 | 16.5 | 9.2 |
| 230.6 ± 0.3 | 0.60 ± 0.03 | 10 | 2 | 13.2 ± 0.7 | 16.2 | 9.0 |
| 83.9 ± 0.6 | 0.80 ± 0.03 | 7 | 1 | 9.8 ± 0.7 | 17.2 | 5.5 |
|  |  |  | 2 | 8.8 ± 1.3 | 77.5 | 46.8 |
| 162.2 ± 1.0 | 0.80 ± 0.03 | 7 | 1 | 11.3 ± 0.7 | 17.2 | 5.5 |
|  |  |  | 2 | 9.6 ± 0.7 | 40.8 | 23.6 |

An analysis of the literature data demonstrated that the existing mathematical models and computations are based on some significant simplifications (in particular, neglect of finiteness of the cell size), and they prove to be unsuitable for comparison with our experimental data. For this reason, an analytical model of displacement in a Hele-Shaw cell, which is suitable for quantitative comparison with experiment, was constructed further in the Dissertation.

A slow quasistationary displacement of one fluid by another in a Hele-Shaw cell has been considered. The two fluids are assumed to be immiscible and incompressible. The motion is thought of as quasidimensional, and all the characteristics of the flow are averaged over the cell thickness $b$. An arbitrarily small distortion of the interface is written in the form $R_{int} = R + \delta \cos(n\varphi)$, where $R$ is the radius of the unperturbed surface, $\delta$ is the perturbation amplitude, and $n$ is the perturbation mode. The mathematical statement of the problem (see also Fig. 4) is as follows:

$$\Delta p_1 = 0, \; \Delta p_2 = 0, \qquad (13)$$

$$M_1 \vec{\nabla} p_1 \cdot \vec{e}\Big|_{R_0} = Q/(2\pi R_0), \; M_1 \vec{\nabla} p_1 \cdot \vec{e}\Big|_{R_{int}} = M_2 \vec{\nabla} p_2 \cdot \vec{e}\Big|_{R_{int}}, \qquad (14)$$

$$p_1 - p_2\Big|_{R_{int}} = 2\varepsilon/b + \alpha V_n^\gamma + \beta K, \; p_2\Big|_{R_\infty} = 0, \qquad (15)$$

where $p_i$ is the fluid pressure ($i = 1$ and 2 for the displacing and the displaced fluid, respectively); $M_i = b^2/12\mu_i$; $\mu_i$ is the fluid viscosity; $\vec{e}$ is the normal to the surface; $R_0$ is the radius of the hole through which the displacing fluid is injected at a constant flow rate $Q$, mm²/s; $\varepsilon$ is the surface tension; $K$ is the interface curvature; $\alpha$, $\beta$ and $\gamma$ are some parameters: $\alpha = 3.8 \cdot 2\varepsilon/b \left(\mu_2/\varepsilon\right)^\gamma$, $\beta = \pi\varepsilon/4$, and $\gamma = 2/3$ (Park C., 1984).

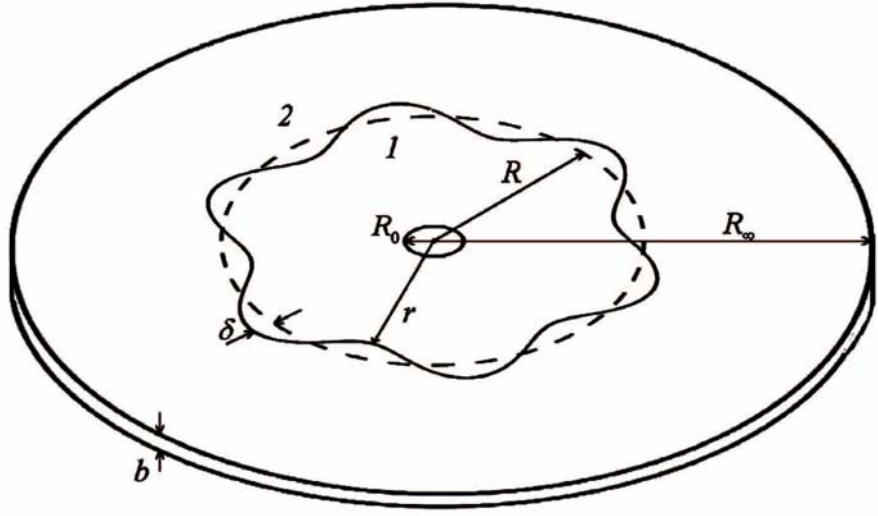

Fig. 4. Radial displacement in the Hele-Shaw cell.

The solution of (13)-(15) in a linear approximation gives the following dependence of the perturbation growth rate:

$$\frac{\dot{\delta}/\delta}{\dot{R}/R}=-1+n\left(1-\frac{M_2}{M_1}\right)\frac{1-\left((n^2-1)\frac{\beta}{R}-\alpha\,\gamma\left(\frac{Q}{2\pi R}\right)^\gamma\right)\frac{2\pi}{Q}\frac{M_1 M_2}{M_1-M_2}}{\frac{M_2}{M_1}\frac{1+(R_0/R)^{2n}}{1-(R_0/R)^{2n}}+\frac{1-(R/R_\infty)^{2n}}{1+(R/R_\infty)^{2n}}+n\alpha\,\gamma\left(\frac{Q}{2\pi R}\right)^\gamma\frac{2\pi M_2}{Q}}. \quad (16)$$

The critical size of the interface stability $R_S$, beyond which the perturbation growth rate $\dot{\delta}$ reverses sign from negative (perturbation damping) to positive (perturbation growth), can be determined from the equation (16) by equating it to zero and solving for $R$. This cannot be done analytically. A numerical analysis of the solutions to (16) is given in the Dissertation. A comparison of the critical radii calculated from (16) and those measured in experiment (see Table 2) shows that slightly smaller values of the critical size were observed in experiment. This distinction can be due to the fact that a linear analysis gives the critical size relative to infinitely small perturbations (a spinodal), and in experiment the perturbations had a finite value. The performed linear analysis for stability also fails to describe the phenomenon of coexistence observed in experiment and does not answer the question how the size of stability will change with increasing amplitude of perturbations. These problems have been solved in terms of the approach based on calculations of the entropy production, which is developed in the Dissertation.

A simple approximation was obtained in the Dissertation for calculating the entropy production for a fluid moving in a Hele-Shaw cell, which is proportional to the radial fluid velocity squared. The approximation is based on an exact expression for the density of the entropy production (as the product of the pressure and velocity-gradient tensors) in a viscous isotropic incompressible fluid with isothermal vortex-free motion. The entropy production was determined when a fluid with a round and a distorted (the function $\delta\cos(n\varphi)$) interface was displaced in the cell. The critical size $R_b$, at which the entropy productions in both displacement regimes are equal, is calculated from the equation

$$\left(\frac{Q}{2\pi R}\right)^2 + \frac{Q}{\pi} M_2 R^{n-1} a_2 n \left(1 + (R_\infty/R)^{2n}\right) = 0, \qquad (17)$$

where $a_2$ is a factor depending on the parameters of the problem (it has a complicated form and is given in the Dissertation). In accordance with the developed approach, the solution of (17) for $R$ allows determining the binodal of the morphological transition from a stable to an unstable interface.

The values of binodals $R_b$ for the parameters corresponding to experimental ones are listed in Table 2. The binodals found for the harmonics, at which the transition *begins*, are smaller than the experimental size. That is, the transition, which was revealed in experiment, occurred in the metastable region.

The expressions (16) and (17) were used in the Dissertation to analyze the metastable regions depending on the basic parameters of the problem. The dependences of the binodal and the spinodal on the problem parameters are similar. The binodal is always smaller than the spinodal at one and the same harmonic. However, spinodals and binodals, which are related to different perturbing harmonics, can intersect (this explains the phenomenon of coexistence). The higher is the viscosity of the displacing fluid, the larger is the number of the intersections (Fig.5).

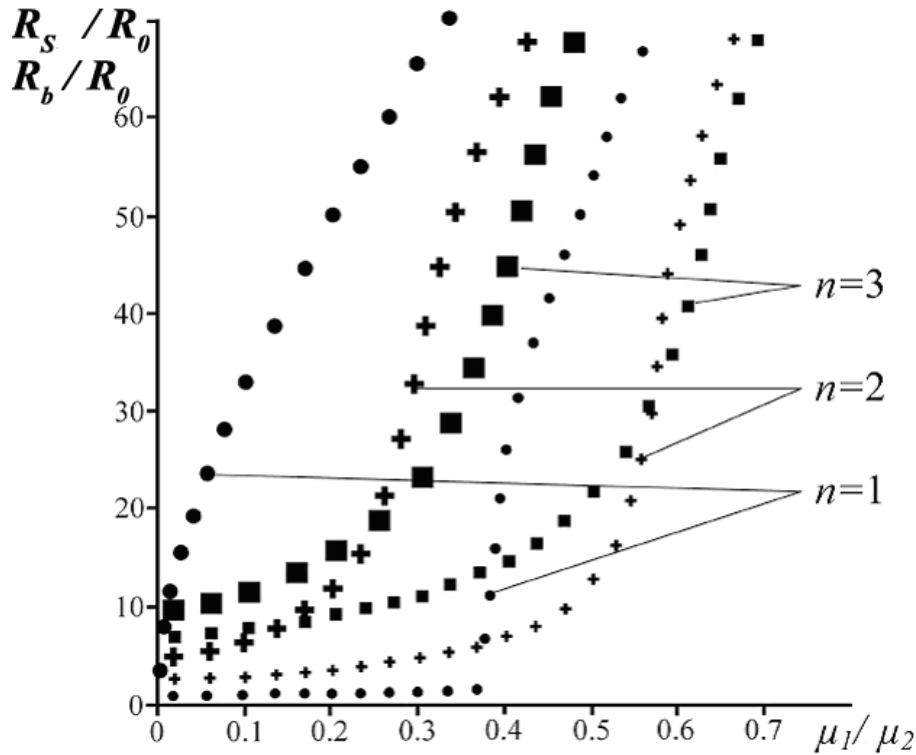

Fig. 5. The dependence of $R_S/R_0$ and $R_b/R_0$ on $\mu_1/\mu_2$. The curves were plotted taking $Q \cdot b = 0.4\ ml/s$, $R_\infty = 20\ cm$, $R_0 = 2\ mm$, $b = 0.6\ mm$, $\varepsilon = 33 \cdot 10^{-3}\ N/m$ and $\mu_2 = 4.65 \cdot 10^{-3}\ kg/(m \cdot s)$. The spinodal and binodal radii are marked with large and small symbols respectively.

In shear experiments the size of stability relative to perturbations with *n* = 1 shifted to the radius of the hole (5.5 mm). According to calculations by (17), it is just 5.5 mm that will be the size of the binodal for the mode with *n* = 1. This fact supports the hypothesis as to the presence of metastable regions and the method of their calculation in the case of displacements in a Hele-Shaw cell. For a more reliable substantiation of this statement, in future it is necessary to compare the boundaries of the metastable regions observed in experiment and predicted theoretically upon introduction of perturbations corresponding to harmonics with *n* > 1.

## BASIC RESULTS

On the strength of the investigations performed for the Dissertation, it is possible to formulate a new area of research: **the maximum entropy production principle as a criterion for selection of possible nonequilibrium (primarily morphological) transitions.** The basic results obtained in this area are as follows.

1. From the analysis of scattered theoretical and experimental studies it was possible to propose a generalized formulation of the maximum entropy production principle (MEPP) in the form: **at each level of description, with preset external constraints, the relationship between the cause and the response of a nonequilibrium system is established such as to maximize the entropy production**.

2. New thermodynamic arguments have been adduced to substantiate MEPP. A relationship between MEPP and the second law of thermodynamics and the minimum entropy production principle has been demonstrated. It has been shown that generalization of the minimum entropy production principle to the integral case is erroneous.

3. The experimentally observed phenomenon of coexistence of morphological phases during nonequilibrium crystallization has been explained by metastability of crystallization regimes.

4. It was proved by analytical and numerical methods that metastable regions limited by a spinodal (the size of absolute instability) and a binodal (the size of absolute stability) exist for simple morphological transitions during nonequilibrium crystallization.

5. Considering an agreement between analytical and numerical calculations and their correspondence to the experimental results, it has been inferred that the maximum entropy production principle allows finding the binodal of a morphological transition during nonequilibrium crystallization. Thus, it has been demonstrated that the entropy production is decisive for description of these nonequilibrium transitions.

6. Kinetic relaxation dependences, which are observed during mass crystallization, have been discussed in the context of the maximum entropy production principle. It was hypothesized that this approach can be extended to other nonequilibrium processes.

7. MEPP was used to predict the smallest Reynolds number of 1200, at which a transition from a laminar to a turbulent flow in a round pipe is possible in the presence of arbitrary perturbations. Experimental facts and calculated values in support of this result have been presented.

8. A translational mechanism of the loss of morphological stability during radial displacement of a fluid by air in a Hele-Shaw cell has been detected in experiment. This fact and a quantitative comparison of the experimental values and the theoretical values predicted in the context of the linear perturbation theory suggest that the theory, which takes into account finiteness of sizes of a radial Hele-Shaw cell, is most suitable for description of experiments.

9. MEPP was used to predict possible sequences of morphological transitions during displacement of a fluid in a Hele-Shaw cell. They allow explaining the coexistence phenomenon and the experimental values of the size of stability upon morphological transitions in a cell.

# BASIC PAPERS ON THE TOPIC OF THE DISSERTATION[9]

---

[9] Martyushev L. M. has also appeared under the alternate (French) spelling Martiouchev L.M.